\documentclass[preprint2]{aastex6}
\usepackage{epsfig}
\usepackage{amsfonts,amssymb,amsmath}

\begin{document}

\title{The Roles of Mass and Environment in the Quenching of Galaxies. II.}
\author{E. Contini$^{1,2}$, Q. Gu$^{1,2}$, X. Ge$^{1,2}$, , J. Rhee$^{3}$, S.K. Yi$^{3}$, X. Kang$^{4}$}

\affil{$^1$School of Astronomy and Space Science, Nanjing University, Nanjing 210093, China; {\color{blue} emanuele.contini82@gmail.com, qsgu@nju.edu.cn}}
\affil{$^2$Key Laboratory of Modern Astronomy and Astrophysics (Nanjing University), Ministry of Education, China}
\affil{$^3$Department of Astronomy and Yonsei University Observatory, Yonsei University, Yonsei-ro 50, Seoul 03722, Republic of Korea}
\affil{$^4$Purple Mountain Observatory, the Partner Group of MPI f\"ur Astronomie, 2 West Beijing Road, Nanjing 210008, China}

\email{emanuele.contini82@gmail.com}
\email{qsgu@nju.edu.cn}

\begin{abstract} 
We take advantage of an analytic model of galaxy formation coupled to the merger tree of an N-body simulation to study the roles of 
environment and stellar mass in the quenching of galaxies. The model has been originally set in order to provide the observed evolution 
of the stellar mass function as well as reasonable predictions of the star formation rate-stellar mass relation, from high redshift to
the present time. 
We analyse the stellar mass and environmental quenching efficiencies and their dependence on stellar mass, halo mass (taken as a proxy
for the environment) and redshift. Our analysis shows that the two quenching efficiencies are redshift, stellar and halo mass dependent,
and that the halo mass is also a good proxy for the environment.
The environmental quenching increases with decreasing redshift and is inefficient below $\log M_* \sim 9.5$, reaches the maximum value 
at $\log M_* \sim 10.5$, and decreases again, becoming poorly efficient at very high stellar mass ($\log M_* \gtrsim 11.5$). 
Central and satellites galaxies are mass quenched differently: for the former, the quenching efficiency depends very weakly on redshift,
but strongly on stellar mass; for the latter, it strongly depends on both stellar mass and redshift in the range $10\lesssim \log M_* \lesssim 11$.
According to the most recent observational results, we find that the two quenching efficiencies are not separable: intermediate mass galaxies are 
environmental quenched faster, as well as intermediate/massive galaxies in more massive haloes.  At stellar masses lower than $\log M_* \lesssim 9.5$ 
both quenching mechanisms become inefficient, independently of the redshift.
\end{abstract}

\keywords{
galaxies: evolution - galaxy: formation.
}

\section[]{Introduction} 
\label{sec:intro}
During the last decades, the scientific community has spent a considerable amount of effort and resources with the 
purpose of learning more about the physical processes that drive galaxy formation and evolution and, among them, 
those responsible for the quenching of the star formation activity in galaxies, which is fundamentally important. Galaxy 
quenching is, indeed, thought to have a remarkable role in shaping galaxy properties, such as morphology, colors and 
stellar age (\citealt{blanton03,baldry04,kauffmann04,brinchmann04,balogh04,cassata08,muzzin12,muzzin13,pallero19,davies19},
and others). 

Nowadays, we all agree that galaxies can be roughly classified in two main categories according to their main properties:
on one side we have star forming systems, which are galaxies that are forming stars, have blue colors, typically young and 
that show late-type morphologies; on the other side we find quiescent or passive systems, which are galaxies with no or 
little star formation activity, that have red colors, typically old stellar ages and that show early-type morphologies
(\citealt{blanton03,kauffmann03,baldry04,noeske07,gallazzi08,wuyts11,wetzel12,vanderwel14}).
Similarly, the same properties are found to be also dependent on stellar mass and environment in such a way that galaxies 
living in a denser environment or that are more massive, are typically early-type systems, less star forming, redder, with 
old stellar ages and more metal rich (\citealt{dressler80,kauffmann03,kauffmann04,baldry06,weinmann06,bamford09,vonderlinden10,peng10,cooper10}).

It appears clear that physical processes that are related to both the surroundings of a given galaxy and its internal conditions 
play a role in quenching its star formation. We usually adopt the term ``environmental quenching" when we refer to the former,
and ``mass quenching" for the latter (\citealt{peng10}). These two modes of quenching together are responsible for the star formation 
history of galaxies, and, although there is not yet a general consensus, there is evidence that they play different roles at different times 
and stellar masses (\citealt{peng12,muzzin12,darvish16,balogh16,kawinwanichakij17,papovich18,pintoscastro19,contini19b}, and references 
therein). 

The term ``mass quenching'' collects all the internal processes that mainly depend, or are linked to the galaxy mass, such as gas outflows
driven by stellar winds or SNe explosions (\citealt{larson74,dekelsilk86,dallavecchiaschaye08}), or AGN feedback from the central supermassive
black hole (\citealt{croton06,fabian12,fang13,cicone14,bremer18}). On the other hand, the term ``environmental quenching'' collects the 
physical processes that quench galaxies because of their interaction between the surrounding area, such as ram pressure stripping (\citealt{gunngott72}),
starvation (or strangulation, \citealt{larson80}) and harassments (\citealt{moore96}).

What process/es, i.e. environmental or mass driven, dominate the quenching of galaxies is still unclear and under debate. It is generally 
thought that mass quenching dominates in massive galaxies and especially at high redshift, while environmental quenching becomes 
important at redshift $z \lesssim 1$ (e.g., \citealt{peng10}). Moreover, it has become clear that, in order to see their single effects, both quenching 
mechanisms have to be studied separately, that is, trying to separate the contribution of one while studying the effect of the other (see, e.g.,
\citealt{muzzin12,darvish16,pintoscastro19}). Both mechanisms can be studied by looking at how the star formation rate (SFR) or specific SFR (SSFR) 
depend on environment, which can be defined as the halo mass, clustercentric distance or the surrounding density, and on stellar mass. Another direct 
way would be by computing their efficiencies, as defined for example in \cite{vandenbosch08} or \cite{peng10} (but see also \citealt{balogh16,fossati17,darvish16,pintoscastro19}
and others), and in particular on how they depend on stellar mass, halo mass (or any other proxy for the environment) and redshift. 

The two ways together are, in principle, supposed to give more information regarding the relative roles of the two quenching modes. For instance,
for what concerns the environmental quenching, when we look at the dependence of the SFR or SSFR on the environment since redshift $z \sim 1$
(when the environment is believed to play a remarkable role), we find no dependence (see, e.g. \citealt{muzzin12,lagana18,contini19b}, and references 
therein for many other works supporting it), which either means that the environment does not play an important role or it acts so fast such that 
it does not influence the star formation of active galaxies but increases the fraction of quiescent galaxies. Many other studies 
(e.g., \citealt{vandenbosch08,quadri12,omand14,balogh16,nantais16,darvish16,fossati17,jian17,kawinwanichakij17,vanderburg18,papovich18,pintoscastro19}),
which have looked at the efficiencies of the environmental and mass quenching, have been able to quantify the two quenching modes, albeit with
different results in terms of mutual dependence among the two modes, redshift, stellar mass and halo mass dependence. This reasoning highlights 
the importance of a quantity that is able to capture the level of quenching, given both by the environment and stellar mass.

What can add more information about the quenching mechanisms is given by the mutual dependence of the two efficiencies, and by how each of them 
depends on redshift. There are observational results supporting the idea that the two quenching processes are separable, such as in 
\cite{baldry06,vandenbosch08,peng10,quadri12,kovac14,vanderburg18}, and more recent works claiming that they are not, such as 
\cite{balogh16,darvish16,fossati17,kawinwanichakij17,papovich18,pintoscastro19,chartab19}. Similarly for what regards the redshift dependence: the environmental 
(or mass) quenching efficiency is found to be dependent on redshift by several studies (such as \citealt{darvish16,nantais16,fossati17,jian17,pintoscastro19}), 
and not dependent by others (such as \citealt{quadri12,balogh16}).

In \cite{contini19b} (hereafter PapI) we took advantage of an analytic model of galaxy formation that was tuned to match the observed evolution 
of the stellar mass function from redshift $z \sim 2.3$ down to the present time and give a reasonable prediction of the evolution of the SFR-$M_*$
relation in the same redshift range, to investigate galaxy quenching by looking at the SFR and SSFR of star forming and quiescent galaxies as a 
function of both stellar mass and environment. Our goal was mainly to understand the roles of environmental and mass quenching at different redshifts.
We concluded that stellar mass is the main responsible for the quenching of galaxies at any redshift investigated (from $0<z<1.5$), and that of the 
environment is just secondary. However, we pointed out the necessity of a further analysis by looking at the efficiencies of the two quenching modes
which, as explained above, can give much more information and better quantify the roles of mass and environment. The aim of the current paper is to 
address this point, by analysing the two quenching efficiencies and their dependences on stellar mass, halo mass and redshift, in order to shed more 
light and put some constraints on the possible physical mechanisms that quench galaxies, and on their relative importance as a function of time.

The paper is organised as follows. In Section \ref{sec:methods} we briefly describe our model and define the quenching efficiencies. In Section \ref{sec:results} 
we present our analysis which will be fully discussed in Section \ref{sec:discussion}, and in Section \ref{sec:conclusions} we draw our conclusions. Throughout 
this paper we use a standard cosmology, namely: $\Omega_{\lambda}=0.73, \Omega_{m}=0.27, \Omega_{b}=0.044$, $h=0.7$ and $\sigma_{8}=0.81$. Stellar masses are 
computed with the assumption of a \cite{chabrier03} Initial Mass Function (IMF), and all units are $h$ corrected.

\section[]{Methods}  
\label{sec:methods}

In the analysis done in PapI we made use of an analytic model that was previously developed in \cite{contini17a} and later refined in \cite{contini17b}. The model 
ran on the merger tree of an N-Body simulation presented in \cite{kang12}, with the main goals of predicting the observed evolution of the stellar mass function and 
giving a reasonable prediction of the SFR-$M_*$ relation as a function of redshift. For what concerns the topic addressed here, a sufficient description of the model 
can be found in PapI, while we refer the reader to \cite{contini17b} for a detailed explanation. Here we briefly put into words the main characteristics and 
implementations important in the context of this paper.

The main feature of the model is to make use of the subhalo abundance matching technique to populate dark matter haloes with galaxies (e.g., \citealt{vale04}). 
The model reads the merger tree of the N-Body simulation where all the information regarding the merger history of each halo is stored. At one particular redshift, which 
we set to $z_{match}=2.3$, the model is forced to match the observed stellar mass function. At redshifts lower than $z_{match}$, the code sorts new born dark matter haloes and, 
by using the stellar mass-halo mass relation at that particular redshift, it assigns a galaxy to each halo. The two novelties of the implementation rely on the way we treat (1) the merger 
histories of a galaxy, and (2) its star formation history \footnote{It must be noted that the model also considers the formation of the intra-cluster light via stellar 
stripping and galaxy mergers (see \citealt{contini14,contini18,contini19a} for more details).}. The former is given directly by the information stored in the merger tree, 
while the latter is treated with a functional form (see equations 1-3 in PapI) which is different according to the type of the galaxy (central or satellite) and that 
depends on the quenching timescale (function of stellar mass, redshift and type of galaxy). For what concerns the treatment of the star formation history, the philosophy 
of our model is very similar to the so-called ``delayed-then-rapid'' quenching mode suggested by \cite{wetzel13}, where satellites evolve as centrals for a given amount 
of time (delay) after accretion, and then quench fast right after (rapid mode).

\subsection[]{Quenching Efficiencies}  
\label{sec:efficiencies}

In order to quantify the quenching due to the environment and stellar mass, in the following analysis we make use of two quantities: the environmental quenching efficiency 
$\epsilon_{env}$, and the mass quenching efficiency $\epsilon_{mass}$. The two quantities were originally introduced by \cite{peng10} (although a similar approach for the 
environmental quenching efficiency was introduced earlier by \citealt{vandenbosch08}), and later used by other authors (e.g., \citealt{quadri12,kovac14,balogh16,kawinwanichakij17,pintoscastro19})
to quantify the relative role of mass and environment in quenching galaxies, even though with slightly different definitions.

We define the environmental quenching efficiency as follows:
\begin{equation}\label{eqn:env_eff}
 \epsilon_{env}(M_{h},M_{h,0},M_{*},z) = \frac{f_q (M_{h},M_{*},z)-f_q (M_{h,0},M_{*},z)}{1-f_q (M_{h,0},M_{*},z)} , 
\end{equation}
where $f_q (M_{h},M_{*},z)$ is the fraction of quiescent galaxies of mass $M_*$ in haloes of mass $M_h$ at redshift $z$, $f_q (M_{h,0},M_{*},z)$ the same fraction but 
for quiescent galaxies in low-mass haloes ($M_h < 10^{13} M_{\odot}$). Basically, for galaxies in a given environment defined as the halo mass, $\epsilon_{env}$ quantifies the excess 
(with respect to what we will refer to as field environment) of galaxies that are quenched because of physical processes linked to the environment at redshift $z$. In practice, we divide the 
sample of galaxies in centrals and satellites. Centrals are selected as field galaxies, i.e. residing in very low mass haloes ($M_h < 10^{13} M_{\odot}$) with no or a few 
satellites around, while satellites are taken within the virial radius $R_{200}$ of haloes with $M_h > 10^{13} M_{\odot}$. It is worth noting that, among centrals, the brightest 
group/cluster galaxies (i.e. those residing in the center of the main haloes) are not included in the analysis. For this reason, centrals are real ``field/isolated'' galaxies. For the 
sake of simplicity, in the rest of paper we will call them ``centrals" or ''field galaxies".

Similarly, the mass quenching efficiency is defined as
\begin{equation}\label{eqn:mass_eff}
 \epsilon_{mass}(M_{*},M_{*,0},M_{h},z) = \frac{f_q (M_{h},M_{*},z)-f_q (M_{h},M_{*,0},z)}{1-f_q (M_{h},M_{*,0},z)} , 
\end{equation}
where $f_q (M_{h},M_{*},z)$ is the fraction of quiescent galaxies of mass $M_*$ in haloes of mass $M_h$ at redshift $z$, and $f_q (M_{h},M_{*,0},z)$ the fraction of quiescent 
galaxies in the same fixed environment given by $M_h$ at lower stellar mass $M_{*,0}$. Essentially, at a fixed environment, $\epsilon_{mass}$ quantifies the fraction of galaxies 
that have been quenched compared to the star forming fraction at lower stellar mass ($M_{*,0}$). In the literature, the completeness limit at a given redshift is often chosen 
as the reference stellar mass $M_{*,0}$ (see, e.g., \citealt{darvish16,kawinwanichakij17,pintoscastro19}). Given the fact that our completeness limit (at any redshift) is far 
lower than that of the above quoted works, we choose $M_{*,0}$ as the stellar mass at which at least 90\% of the galaxies are star forming. However, as we will see below,
our model overestimates the SFR for low mass galaxies, so our choice translates in $\log M_{*,0} \simeq 9.6-9.7$ for satellites in the redshift range considered (except for 
$z=1.5$ for which $\log M_{*,0} \sim 10.4$), and $\log M_{*,0} \simeq 10.5$ for centrals, similar to $M_{*,0}$ of many observational studies.

There are some caveats concerning the definitions of the two efficiencies that are worth mentioning. The first one is with regard to the definition of the environmental quenching
efficiency $\epsilon_{env}$. The excess of galaxies that are quenched with respect to the field, and in our case the excess of satellites (cluster environment) with respect to
centrals (field) is calculated at the same epoch. As many authors noted (e.g., \citealt{balogh16,vanderburg18}), a better approach would be to consider the quenched
fraction in the field at the time of accretion. Theoretically speaking, this would be possible with models of galaxy formation that are able to trace the history of any single 
galaxy, such as ours, but observationally impossible without assumptions (which anyway would not avoid the progenitor bias issue) on the different galaxy populations. In the 
following analysis, we decide to keep the same observational approach (the classical definition), so to have fair comparisons with observational works. Another important
issue concerning the definition of $\epsilon_{env}$ relies on the fact that it does not consider the differential growth in stellar mass of galaxies, which might not be negligible,
simply because the two populations are taken at the same redshift. 

There is a second issue regarding the definitions of both efficiencies, which makes direct comparisons not simple, and relies on the definition of the environment itself. As noted 
in PapI, the definition of the environment in the literature spans from the mass of the halo where the galaxy resides, to clustercentric distance or local galaxy density. In this 
work, we first separate satellite from central galaxies, which already is a first level of environmental separation, and then, among satellites we define their environment by using 
their halo mass (not to be confused with their subhalo mass) as a proxy.  

Having in mind the discussion above, in the following section we present our analysis where we focus mainly on:
\begin{itemize}
 \item [a)] quantifying the environmental quenching efficiency $\epsilon_{env}$ and its dependence with redshift and halo mass;
 \item [b)] quantifying the mass quenching efficiency $\epsilon_{mass}$ and its dependence with redshift and stellar mass;
 \item [c)] the mutual dependence of the two efficiencies, i.e. whether $\epsilon_{env}$ depends on stellar mass and $\epsilon_{mass}$ depends on the environment.
\end{itemize}
Given the different results found in the recent past, point (c) is particularly interesting. The case of mutual dependence of the two efficiencies would mean that the 
two quenching modes are not separable.

\section{Results}
\label{sec:results}

In this section, we present the main results of our model, and their interpretation will be discussed in Section \ref{sec:discussion}, which will also consider a full comparison 
with the most relevant works. As we need to separate star forming from quiescent galaxies, we decide to follow the same approach used in PapI, i.e. we use an SSFR cut that is redshift dependent.
At a given redshift, we consider star forming all those galaxies with SSFR higher than $t_{hubble}^{-1}$ and quiescent all those with SSFR lower than that. In order to study the effect of 
the environment, we split our sample of galaxies according to their rank, central or satellites. As mentioned above, centrals are isolated galaxies which we consider as the ``field'' 
environment, while satellites are galaxies that reside within the virial radius of groups/clusters with mass higher than $10^{13} M_{\odot}$. For satellites we will consider the effect of 
being part of groups/clusters of different mass since the halo mass is our proxy for the environment.

\begin{figure*}
\includegraphics[scale=0.9]{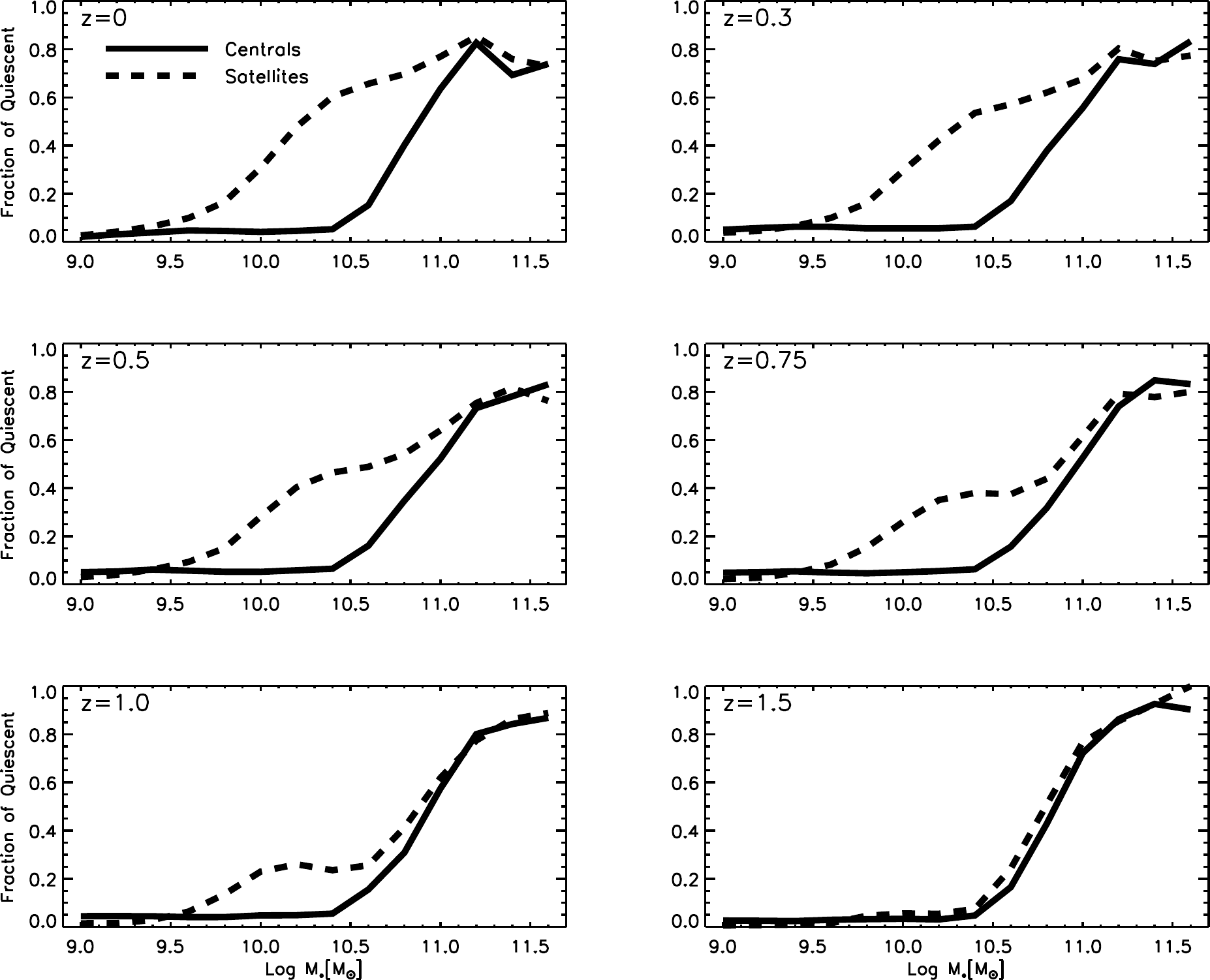}
\centering
\caption{Fraction of quiescent central (solid lines) and satellite (dashed lines) galaxies as a function of stellar mass at different 
redshifts (different panels), from $z=1.5$ to $z=0$. The fraction of quenched satellites is higher than the fraction of quenched centrals at
almost all masses ($\log M_* \simeq [9.5-11]$) and redshifts ($z=[0.0-1.0]$).}
\label{fig:quiescent_fraction}
\end{figure*}

Figure \ref{fig:quiescent_fraction} plots the fraction of quiescent central (solid lines) and satellite (dashed lines) galaxies as a function of stellar mass at different redshifts (different panels), 
from $z=1.5$ to $z=0$. We note that there is a clear dependence with stellar mass for both types of galaxies, being the fraction higher at increasing stellar mass, independent of the reshift. 
Moreover, in the redshift range $z=[0.0-1.0]$ (i.e., with the only exception of $z=1.5$), the fraction of quenched satellites is higher than that of quenched centrals at almost all stellar masses 
($\log M_* \simeq [9.5-11]$), except for low and high mass galaxies. Compared with most of the observational fractions, the predictions of our model are low in the low mass end. As we pointed out in PapI, 
the different definitions for separating star forming from quiescent galaxies can play a significant role, but we also noticed that our model could overestimate the SFR history of low mass galaxies. This problem, 
if real, can bias the efficiency of quenching (both environmental and mass) for galaxy with low stellar mass. Having in mind this caveat, in the following two subsections we present a detailed analysis of the 
quenching modes by studying the dependences of $\epsilon_{env}$ and $\epsilon_{mass}$ on stellar mass, halo mass and redshift.

\subsection{Environmental Quenching Efficiency}
\label{sec:env_quenching}

\begin{figure*} 
\begin{center}
\begin{tabular}{cc}
\includegraphics[scale=.47]{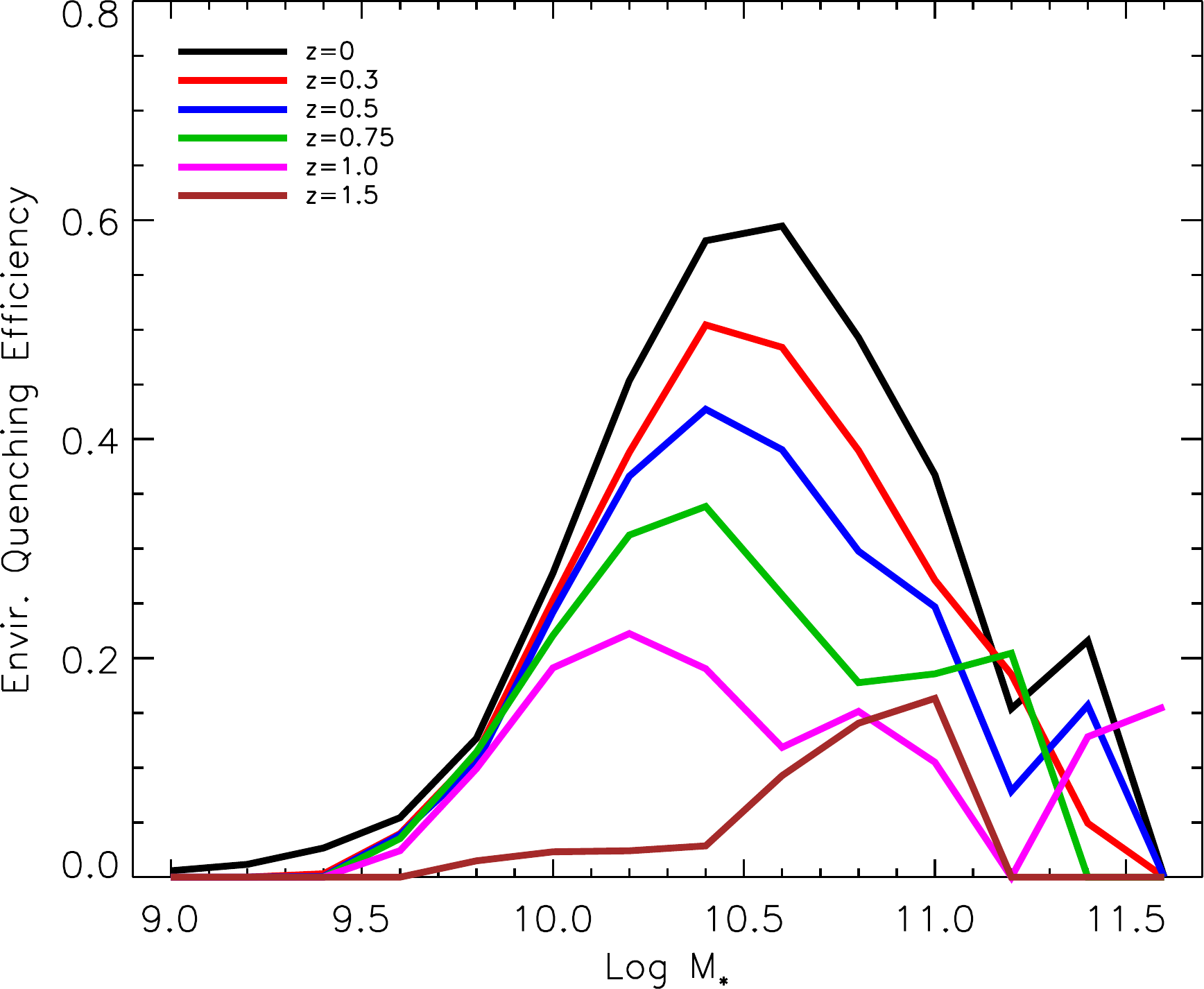} &
\includegraphics[scale=.47]{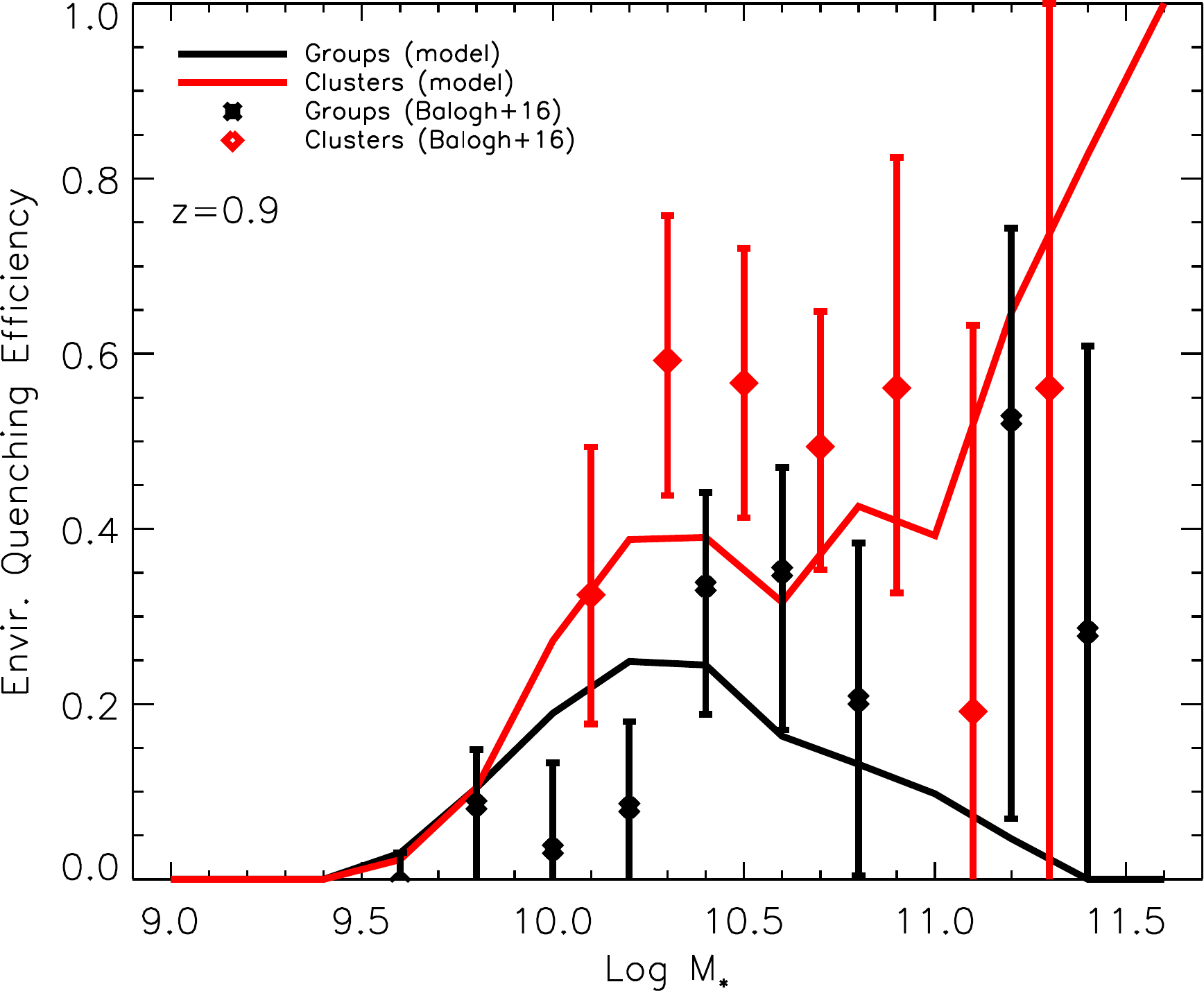} \\
\end{tabular}
\caption{Left panel: environmental quenching efficiency as a function of stellar mass at different redshifts (different colors), from $z=1.5$ to $z=0$.
Right panel: environmental quenching efficiency as a function of stellar mass at $z=0.9$, for groups (black line) and clusters (red line) as predicted by
our model and compared with observed data (black crosses and red diamonds) by \cite{balogh16} in the same redshift ($0.8<z<1.2$) and halo mass ranges 
($13.5<\log M_{halo} <14.0$ for groups, and $\log M_{halo}>14.2$ for clusters).}
\label{fig:fc_mass}
\end{center}
\end{figure*}

\begin{figure*} 
\begin{center}
\begin{tabular}{cc}
\includegraphics[scale=.47]{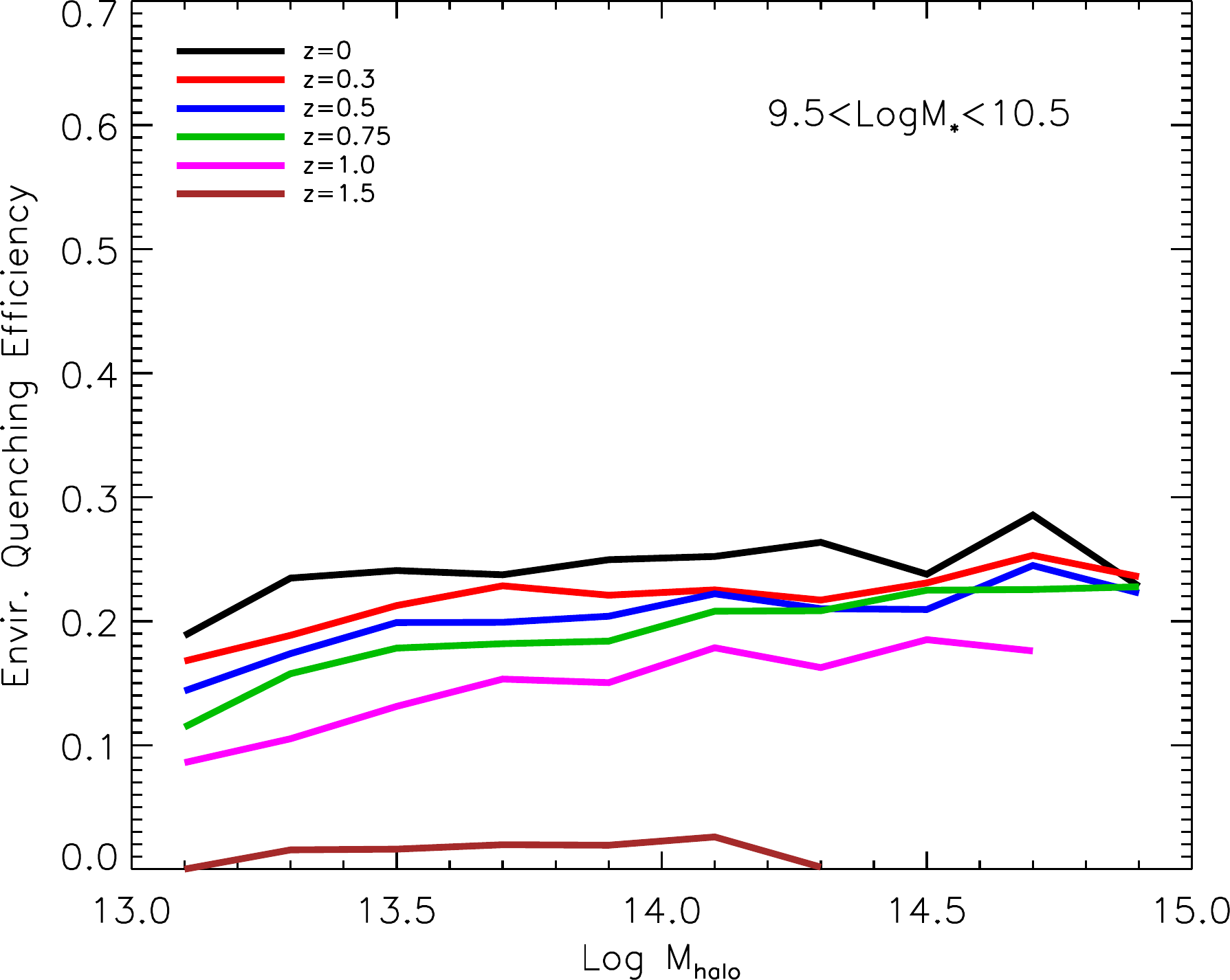} &
\includegraphics[scale=.47]{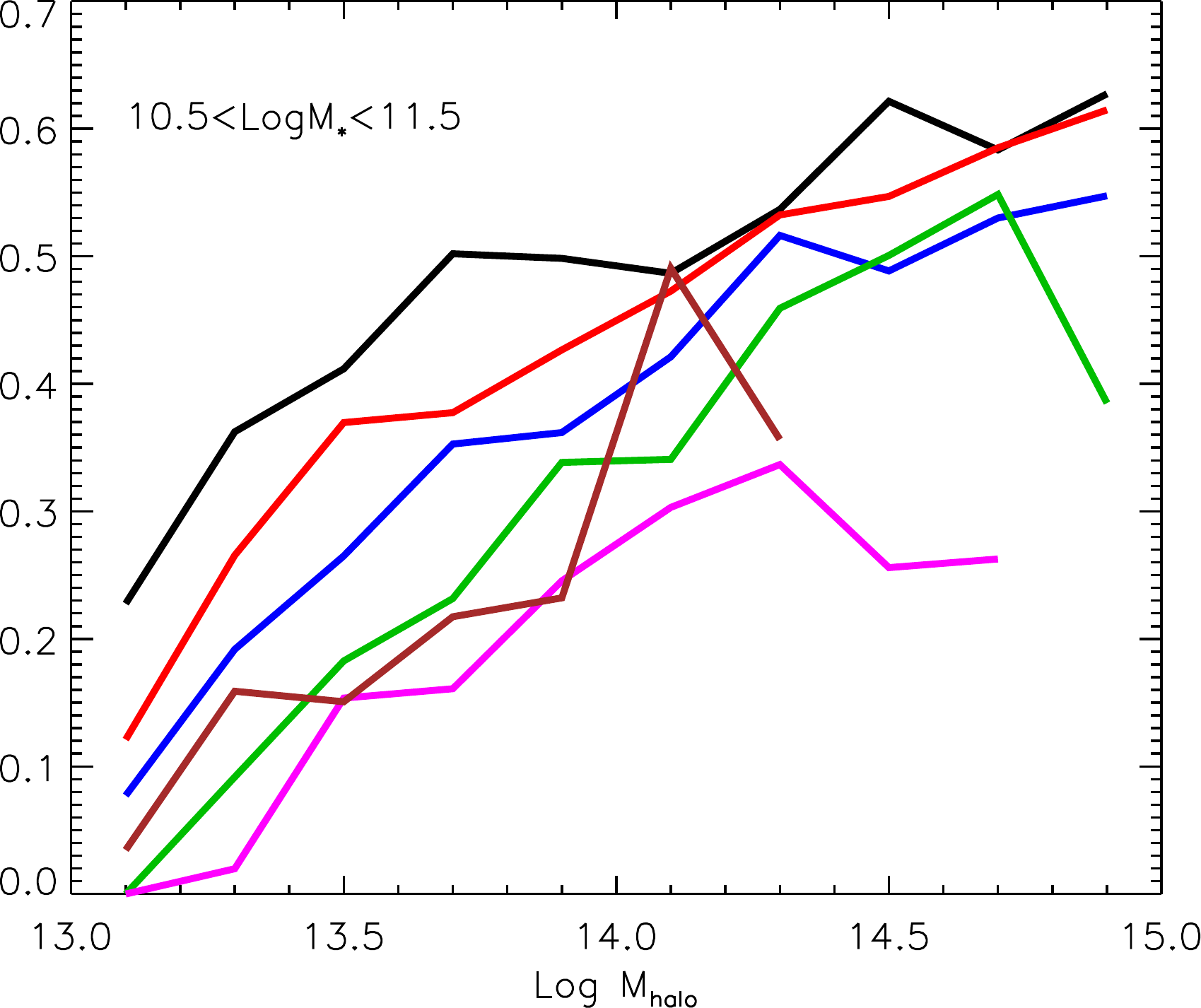} \\
\end{tabular}
\caption{Environmental quenching efficiency as a function of halo mass at different redshifts (different colors), for galaxies in two stellar mass ranges
(different panels). The efficiency of environmental quenching does not depend strongly on the halo mass for low mass galaxies, while it strongly depends on
halo mass for more massive galaxies. In both stellar mass ranges, a redshift dependence is seen, that is, the efficiency is higher with decreasing redshift.}
\label{fig:fc_halomass}
\end{center}
\end{figure*}

\begin{figure*} 
\begin{center}
\begin{tabular}{cc}
\includegraphics[scale=.47]{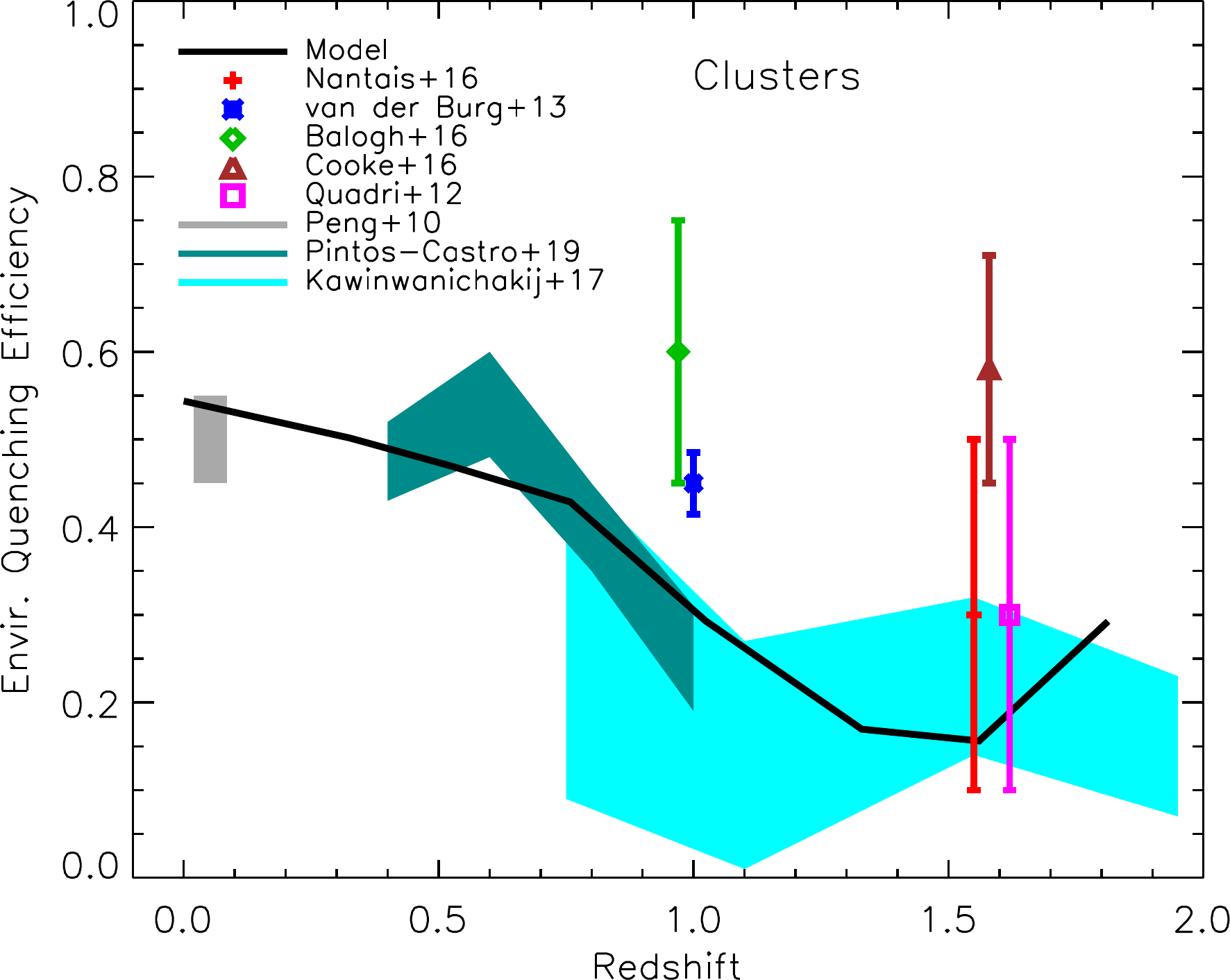} &
\includegraphics[scale=.47]{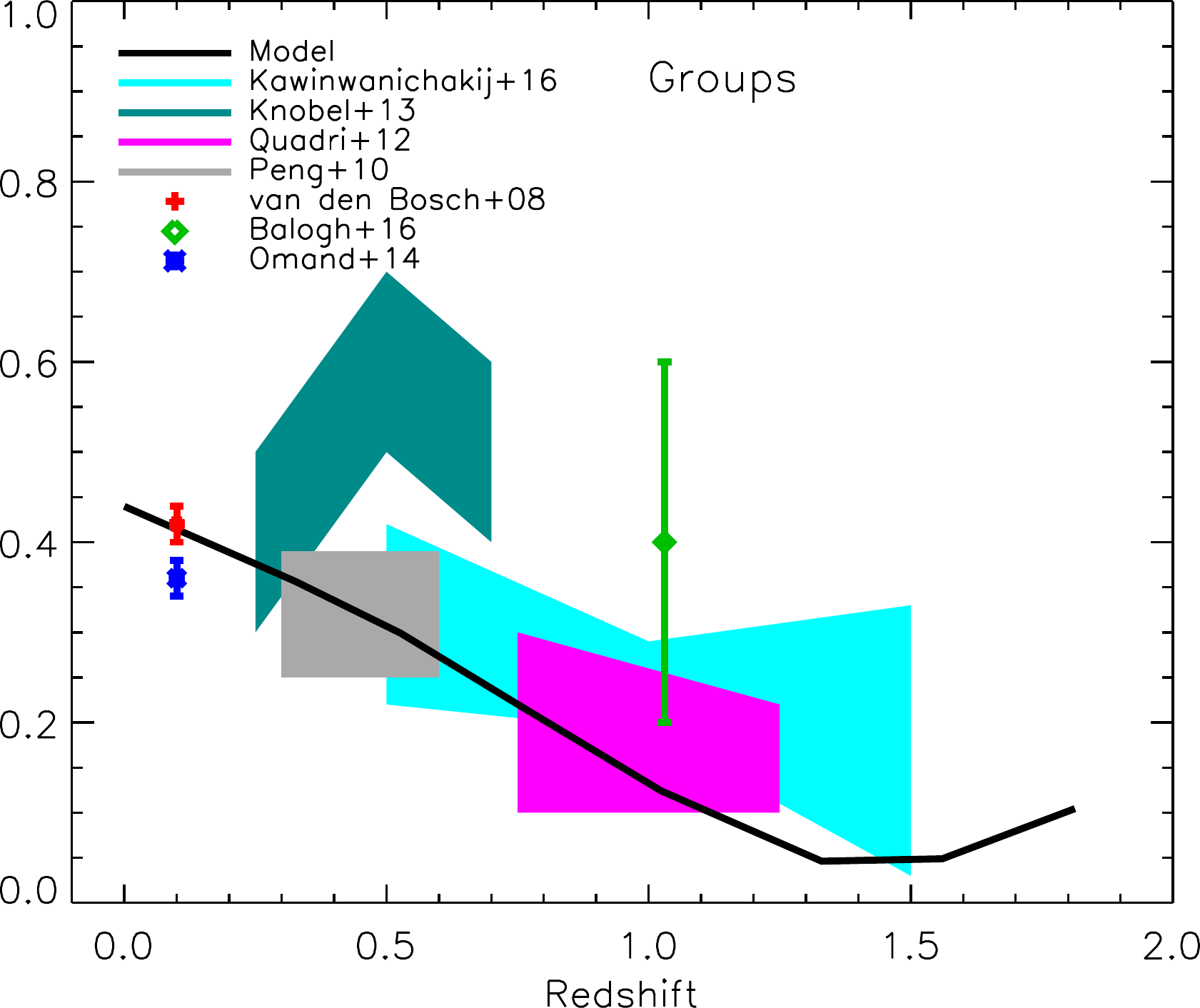} \\
\end{tabular}
\caption{Environmental quenching efficiency as predicted by the model (solid lines) as a function of redshift in clusters (left panel) and groups (right panel), 
compared with several observed data. Overall, our model is able to reproduce the observed trend (lower efficiency with increasing redshift) in both groups and clusters.}
\label{fig:fc_redshift}
\end{center}
\end{figure*}

We make use of Equation \ref{eqn:env_eff} to derive the environmental quenching efficiency $\epsilon_{env}$ and plot it as a function of stellar mass at different 
redshifts in the left panel of Figure \ref{fig:fc_mass}. $\epsilon_{env}$ clearly depends on redshift, being higher with decreasing redshift, and on stellar mass, being 
lower for low and high mass galaxies. The trend with redshift that we find is in agreement with many other works (see, e.g., \citealt{nantais16,darvish16,fossati17}), but 
in disagreement with others (e.g., \citealt{quadri12,balogh16}). However, it must be noted there is a growing consensus for an increasing efficiency with decreasing redshift,
as one would expect if the environment plays a more important role at low redshift. 

The dependence of $\epsilon_{env}$ on stellar mass is a more delicate issue. In the right panel of Figure \ref{fig:fc_mass} we compare the dependence of $\epsilon_{env}$ on 
stellar mass as predicted by our model with observational results by \cite{balogh16}, for groups (black line and crosses) and clusters (red line and diamonds) at redshift $z\sim 0.9$. 
Groups and clusters have been chosen in the same redshift range ($0.8<z<1.2$) and halo mass ranges as in Balogh et al., i.e. $13.5<\log M_{halo} <14.0$ for groups, and 
$\log M_{halo}>14.2$ for clusters. The predictions of our model agree fairly well with the observed data, in particular the decrease of the efficiency at lower stellar mass in groups. 
\cite{balogh16} find that the environmental quenching efficiency decreases with stellar mass and becomes inefficient at $\log M_{*} <9.5$. Their results are supported by \cite{papovich18}, 
who study the effect of the environment in shaping the evolution of the stellar mass function in the redshift range $0.2<z<2.0$, and find that the evolution of the stellar mass function of quiescent 
galaxies implies a decreasing $\epsilon_{env}$ with decreasing stellar mass. Our model is in line with this picture and, considering the caveat discussed above (the overestimation 
of the SFR of low mass galaxies discussed in Section \ref{sec:results}), it is an evidence that our results are safe even at low stellar mass. Moreover, although from the left panel 
of Figure \ref{fig:fc_mass} we see that $\epsilon_{env}$ of high mass galaxies is low compared to intermediate mass galaxies, in the right panel we show that the environmental efficiency 
of high mass galaxies in clusters is higher than that of high mass galaxies in groups. We will come back on, and discuss both points below, where will provide more evidence supporting these results.

Now we study the dependence of the environmental quenching efficiency on the halo mass, as a function of redshift and at fixed stellar mass. This information is shown in the two 
panels of Figure \ref{fig:fc_halomass}, for galaxies with stellar mass in the range $9.5<\log M_* <10.5$ (left panel), and $10.5<\log M_* <11.5$ (right panel). As in the previous 
figure, the efficiency is higher at lower redshift, but there are two points worth noting from the information given by this figure: a) the efficiency does not depend on halo mass 
(i.e. environment) for the lowest stellar mass galaxies and, b) it does in the highest stellar mass galaxies, regardless of the redshift. For instance, while in the left panel the 
efficiency at the present time is basically constant at the value $\sim 0.25$, in the right panel it spans the range 0.2 in low mass haloes to 0.6 in high mass haloes. This is a 
clear evidence that the environment acts differently on galaxies of different mass. To summarise so far, the environmental quenching efficiency is stellar mass dependent and 
also environmental dependent. 

We conclude the analysis concerning the environmental quenching efficiency with Figure \ref{fig:fc_redshift}, which plots the integrated $\epsilon_{env}$ as a function of redshift 
for clusters (left panel), and groups (right panel). Our model (solid black lines) is compared with several observations (different points and shaded regions). For the sake of 
honesty, we must stress that these comparisons are not 100\% fair for three simple reasons: different authors have different stellar mass completeness limit in deriving $\epsilon_{env}$
and the halo mass ranges (group or cluster) are also different (we used the same ranges in halo mass adopted in Figure \ref{fig:fc_mass}). Moreover, the proxy for the environment is 
also different from author to author (e.g., halo mass, local density and clustercentric distance) and yields to different definitions of $\epsilon_{env}$. Given these caveats, we can 
reasonably state that our model predicts the observed trend with redshift and matches most of the observations from high redshift to the present time, in groups and clusters.

In order to highlight the most important results so far we can simply summarise as follows: the environmental quenching efficiency is redshift, stellar mass, and environmental dependent.

\subsection{Mass Quenching Efficiency}
\label{sec:mass_quenching}

\begin{figure*}
\includegraphics[scale=0.9]{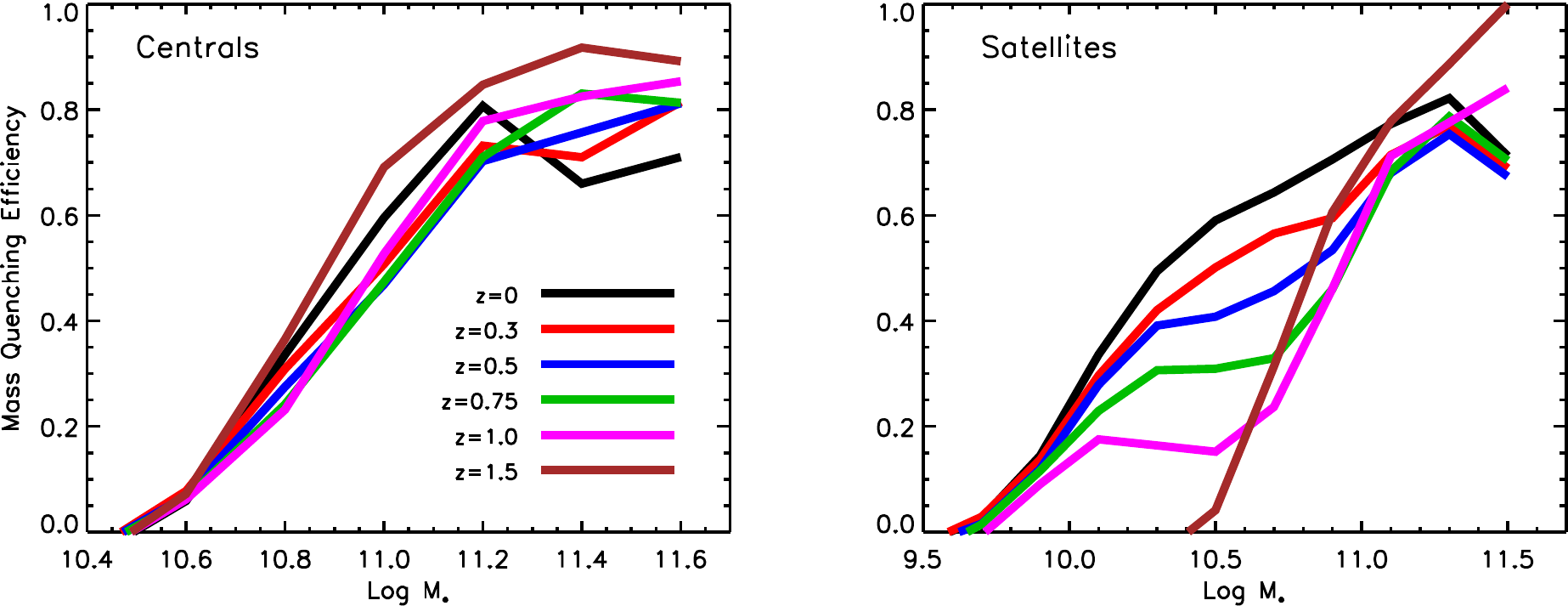}
\centering
\caption{Mass quenching efficiency as a function of stellar mass at different redshifts (different colors), for central (left panel) and satellite (right panel) galaxies.
There is a clear trend with stellar mass, in that more massive galaxies are more efficiently quenched by internal processes, for both centrals and satellites. Moreover, although 
the dependence on redshift is not clear for centrals, satellites in the intermediate-massive stellar mass range are more efficiently quenched at lower redshift.}
\label{fig:mqe_mass1}
\end{figure*}

\begin{figure*}
\includegraphics[scale=0.9]{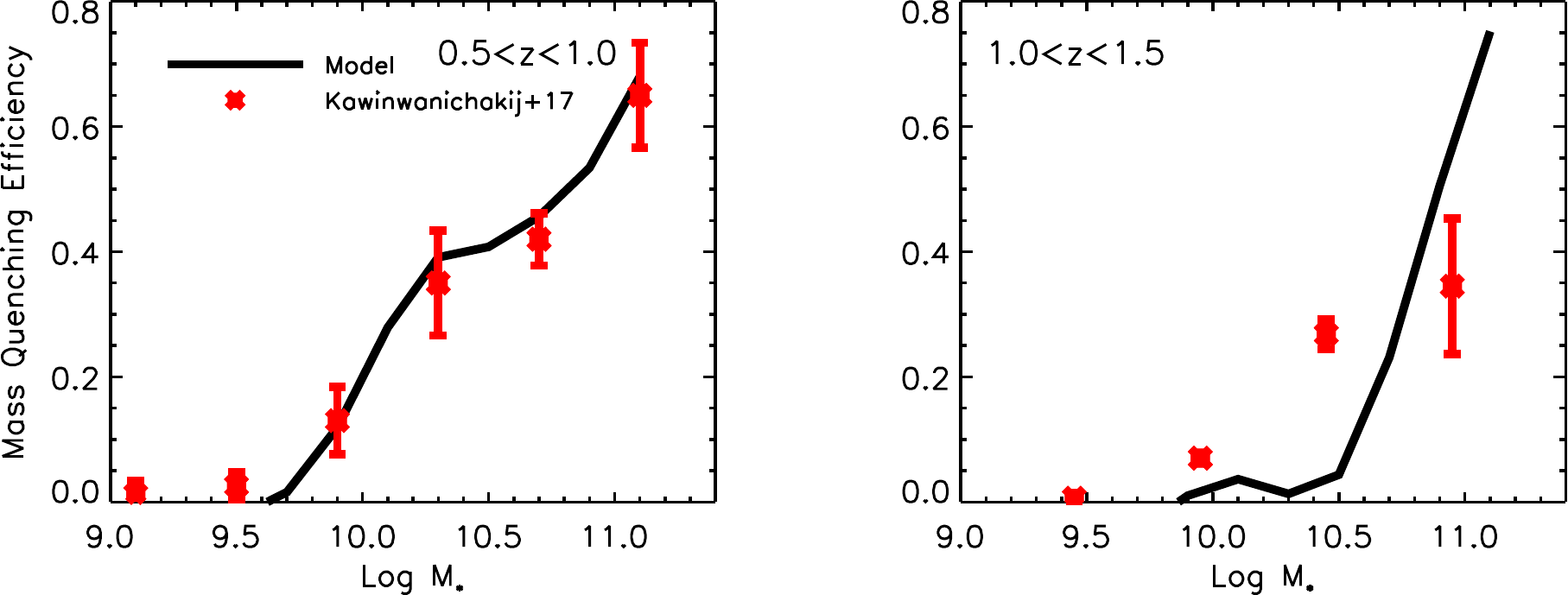}
\centering
\caption{Comparison of the predictions of our model with the observed data by \cite{kawinwanichakij17} for the mass quenching efficiency as a function of stellar mass in 
two redshift ranges (different panels), $0.5<z<1.0$ and $1.0<z<1.5$. Our model matches very well the observed data in the lowest redshift range, and although it captures 
the overall trend with stellar mass even at higher redshift (increasing efficiency with increasing stellar mass), it does not match the observed data. In order to be as 
consistent as possible with the observed data (galaxies in the highest density environments of \citealt{kawinwanichakij17}'s sample), the predictions of the model consider only 
satellite galaxies.}
\label{fig:mqe_mass2}
\end{figure*}

\begin{figure*} 
\begin{center}
\begin{tabular}{cc}
\includegraphics[scale=.47]{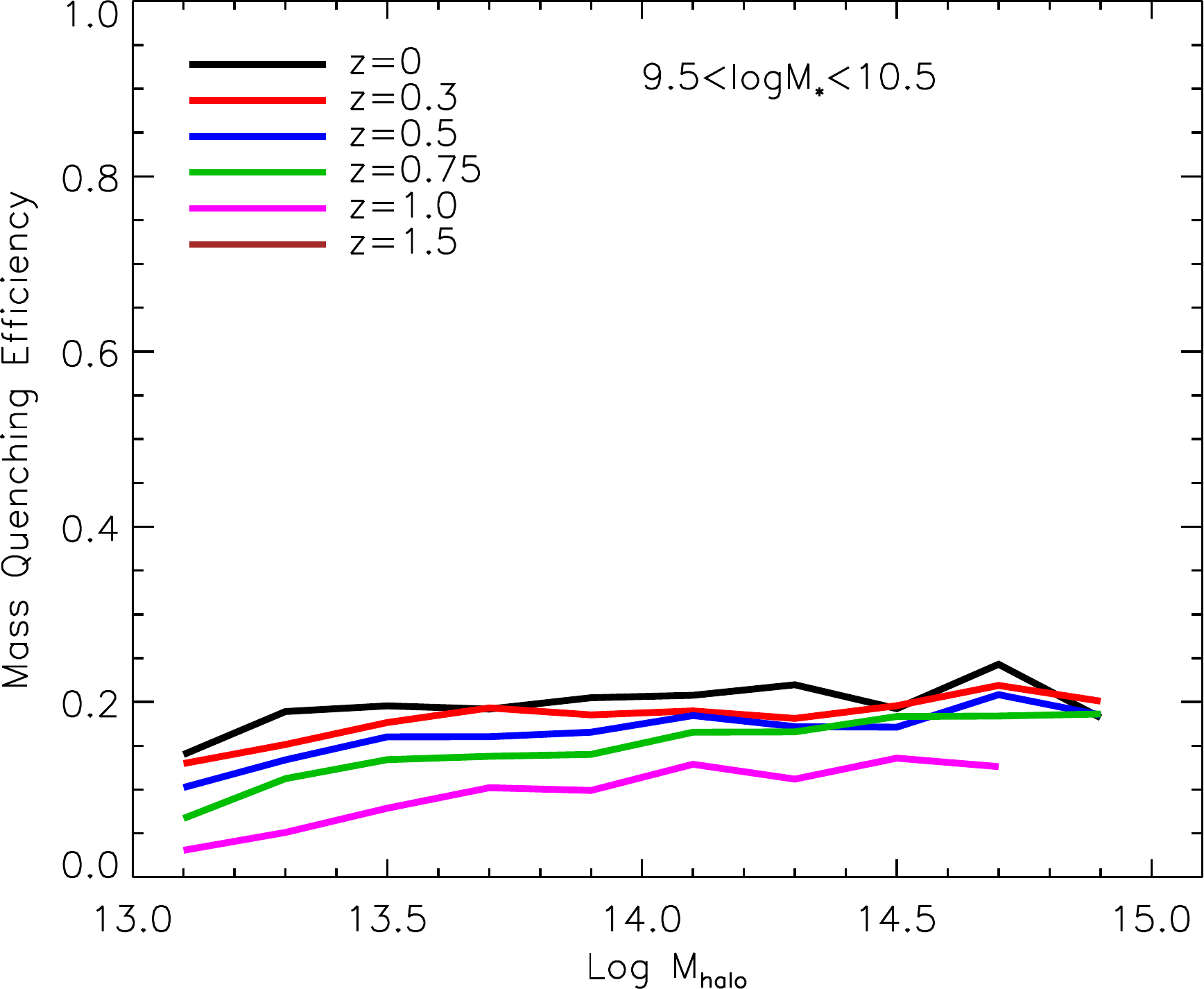} &
\includegraphics[scale=.47]{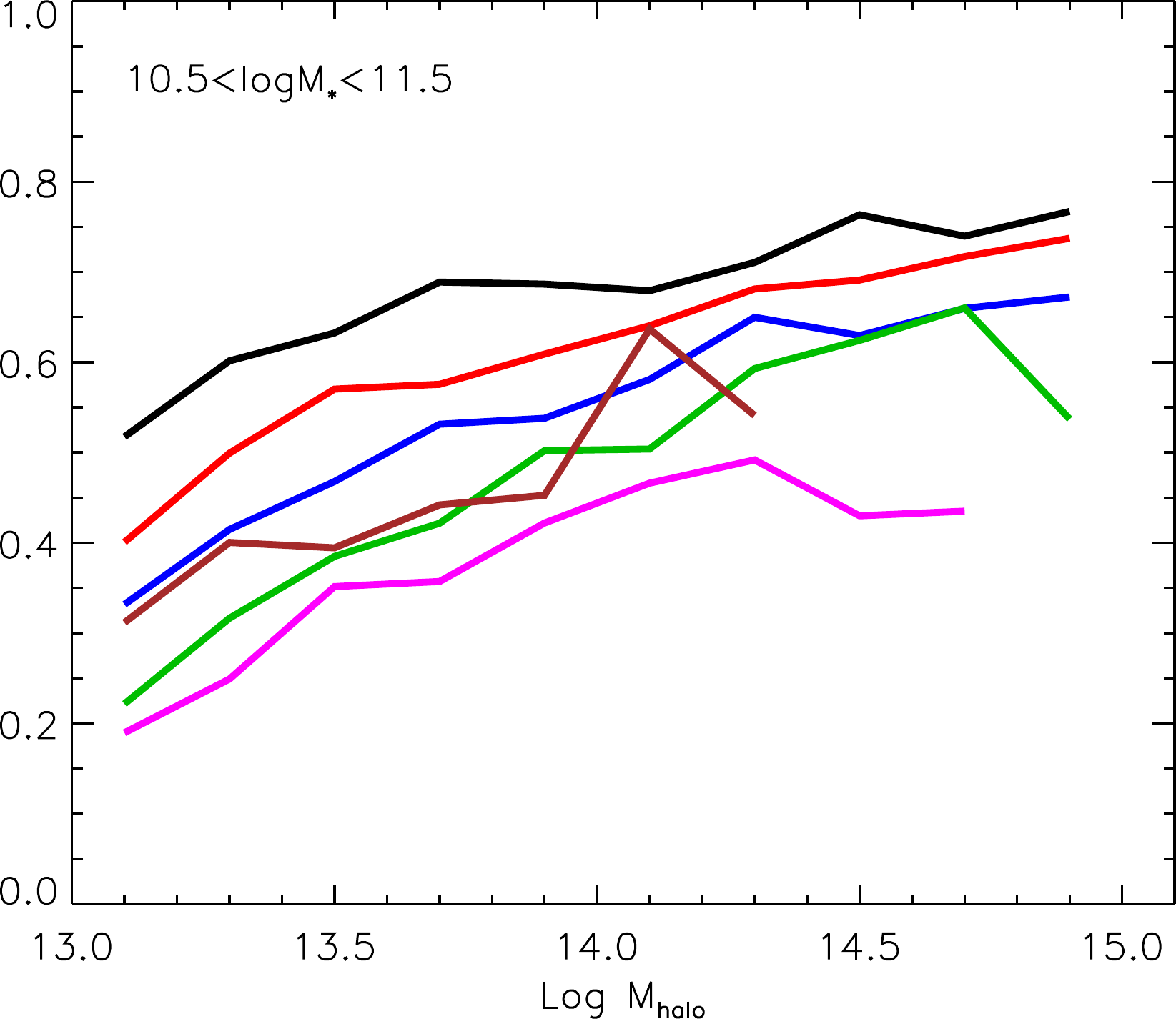} \\
\end{tabular}
\caption{Mass quenching efficiency as a function of halo mass at different redshifts (different colors), for galaxies in two stellar mass ranges
(different panels). The efficiency of mass quenching does not depend strongly on the halo mass for low mass galaxies, while it does depend on
halo mass for more massive galaxies. Similarly to Figure \ref{fig:fc_halomass} (environmental quenching efficiency as a function of halo mass), the efficiency 
is redshift dependent, being higher with decreasing redshift.}
\label{fig:mqe_halomass}
\end{center}
\end{figure*}

\begin{figure*}
\includegraphics[scale=0.9]{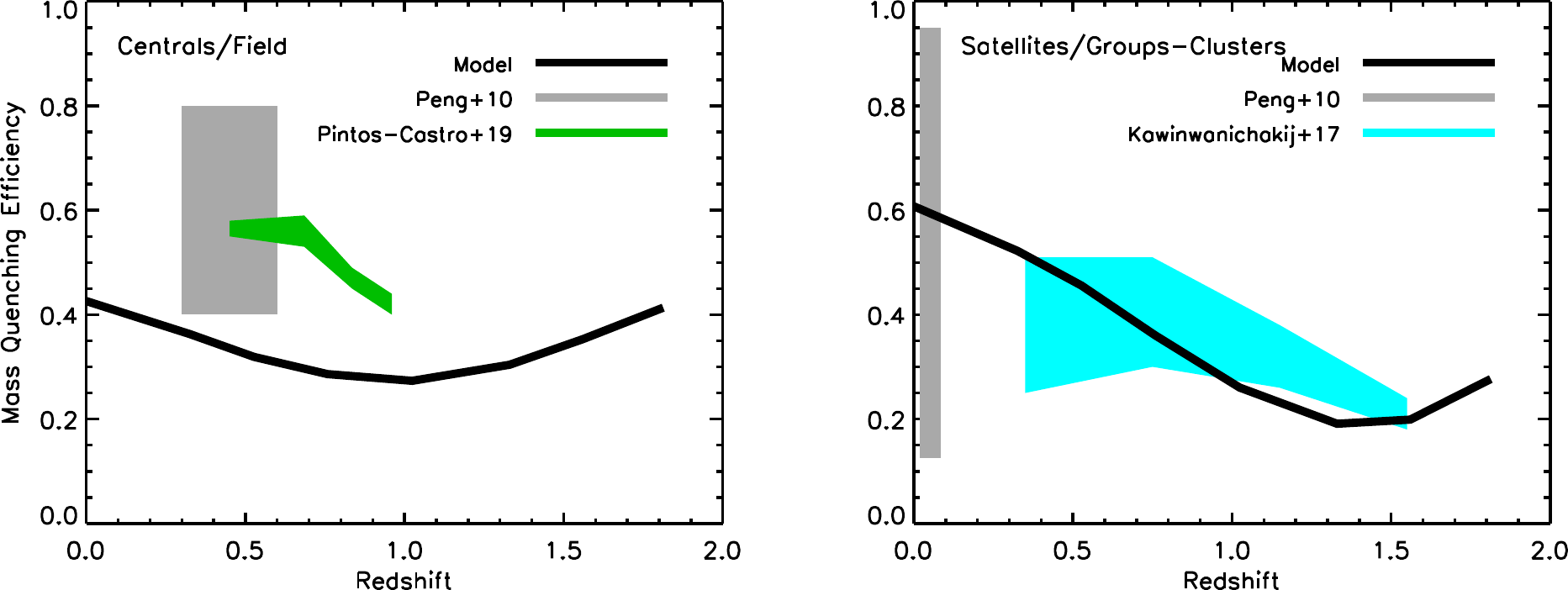}
\centering
\caption{Mass quenching efficiency as predicted by the model (solid lines) as a function of redshift for central/field (left panel) and satellite/group-cluster galaxies (right panel), 
compared with several observed data. The results reflect those found in Figure \ref{fig:mqe_mass1} for both centrals (no clear trend with redshift) and satellites (decreasing efficiency 
with decreasing redshift). The model slightly underpredicts the mass quenching efficiency for central/field galaxies, while it captures the trend and matches the observed data for 
satellite/group-cluster galaxies.}
\label{fig:mqe_redshift}
\end{figure*}

Analogously to the analysis done in the previous subsection, here we make use of Equation \ref{eqn:mass_eff} to derive the mass quenching efficiency, and study its dependence on redshift,
stellar and halo mass. In Figure \ref{fig:mqe_mass1} we show $\epsilon_{mass}$ as a function of stellar mass for central (left panel) and satellite (right panel) galaxies, at different 
redshift as indicated in the legend \footnote{The stellar mass ranges in the two panels are different because central galaxies below $\log M_* \sim 10.4$, and satellite galaxies below 
$\log M_* \sim 9.5$ are all star forming.}. The plots show a clear dependence of $\epsilon_{mass}$ on stellar mass for both centrals and satellites, being more massive galaxies more efficiently 
mass quenched, regardless the redshift. Another important information comes when comparing the two panels: while for central galaxies the dependence of $\epsilon_{mass}$ with redshift is 
not clear, although it is stronger at higher redshift for very massive centrals ($\log M_* \gtrsim 11.3$), satellites in the intermediate/massive stellar mass range ($10 \lesssim \log M_* \lesssim 11$) 
are more efficiently quenched at lower redshift, while the trend reverts to that of high mass centrals in the high mass range \footnote{It must be noted that the predictions at very high stellar 
mass and at high redshift have to be taken with caution due to the poor statistics, especially for satellite galaxies.}. These results are qualitatively in agreement with several observations 
(e.g., \citealt{darvish16,kawinwanichakij17}).

In order to quantify the accuracy of our model in predicting the observed relation between $\epsilon_{mass}$ and stellar mass at different redshifts, in Figure \ref{fig:mqe_mass2} we show 
the predictions of our model (solid black lines) compared with the observed data by \cite{kawinwanichakij17} (red crosses) in the redshift ranges $0.5<z<1.0$ (left panel) and $1.0<z<1.5$
(right panel). In the lowest redshift range our model matches perfectly the observed data, while in the highest redshift range, although qualitatively it captures the observed trend, it 
fails quantitatively.

We now want to quantify the role of the environment on the mass quenching efficiency by plotting $\epsilon_{mass}$ as a function of halo mass. This is shown in Figure \ref{fig:mqe_halomass}, 
at different redshifts (different colors), and for galaxies in two stellar mass ranges (different panels), as shown in the legend. Analogously to Figure \ref{fig:fc_halomass} for $\epsilon_{env}$,
we find that the mass quenching efficiency does not strongly depend on halo mass and only weakly on redshift for the lowest stellar mass range, while it clearly depends on both in the highest stellar 
mass range. Then, this means that the environment (meant as the halo mass) has a strong influence on the mass quenching, another evidence that the two quenching modes are not separable. Moreover, 
by comparing figures \ref{fig:fc_halomass} and \ref{fig:mqe_halomass}, we can note by eye that, for massive galaxies and regardless of the redshift, $\epsilon_{mass}$ is always larger than $\epsilon_{env}$,
in good agreement with the observational results by \cite{darvish16}. 

To complete our analysis on $\epsilon_{mass}$, we show in Figure \ref{fig:mqe_redshift} the integrated mass quenching efficiency, i.e. the total $\epsilon_{mass}$ for galaxies with $\log M* >10.2$
(in order to have a fair comparison with the observed data shown), as a function of redshift, for central/field (left panel) and satellite/group-cluster (right panel) galaxies. The predictions of our model 
(solid lines) are compared with the observed data in the literature. The integrated mass quenching efficiency does not depend much on redshift in the case of central/field galaxies (which reflects 
the result found in Figure \ref{fig:mqe_mass1} for centrals), and our model underestimates the observed data available (\citealt{peng10,pintoscastro19}). The stellar mass ranges on which the 
mass efficiency is calculated are comparable but not exactly the same \footnote{For instance, the stellar mass range considered by \citealt{pintoscastro19} is $10.2< \log M_* <10.8$, while ours is
$\log M_* > 10.5$ because all centrals below that stellar mass are star-forming in our model.}, which might be, at least in part, one possible reason for the mismatch. Other two possible reasons can be 
the different definitions of field/central galaxies and the separation between star forming and quiescent galaxies. On the other hand, the model is able to 
catch the observed trend for satellite/group-cluster galaxies, and to match the available observed data. Moreover, while centrals are mass quenched with roughly the same efficiency, around 0.4, 
satellites experience different mass quenching efficiencies with time, from around 0.2 at $z\sim 1.5$ to 0.6 at the present time, which means that $\epsilon_{mass}$ is a factor 3 times higher at $z=0$ than it is 
at high redshifts. We will come back on this point in Section \ref{sec:epsilon_mass}.

To summarise the most important results of this subsection, we can state that the mass quenching efficiency is redshift, stellar mass, and environmental dependent. 

\section{Discussion}
\label{sec:discussion}
The aim of this work is to extend the results found in PapI by looking at the two quenching modes, due to internal processes that are linked to the stellar mass of a galaxy (mass quenching), and 
due to physical processes that are related to the surroundings of a galaxy (environmental quenching). These two processes together are believed to be responsible for the star formation history of 
galaxies, and generally to act with different timescales. Quantifying their roles becomes then necessary if we want to understand which process/es are important and at what time for a galaxy with a 
given stellar mass. In the literature there is not general consensus regarding what mode dominates and at which time, mainly because different authors look at different galaxy properties which 
give opposing results. 

In PapI we studied the roles of mass and environment in quenching galaxies by analysing the SFR and SSFR as a function of stellar mass, halo mass and redshift. According to our analysis, the main 
conclusion was that mass quenching dominates at any redshift, while the environment plays only a marginal role. We stated that all the results could have been put together logically if environmental 
processes act very fast (\citealt{jung18}) in a way that they do not affect the star formation activity, but can increase the probability of galaxies to become quiescent. In the last decade many authors have studied 
mass and environmental quenching by looking at their efficiencies (with all the caveats discussed above). The main pro of following such an approach comes from the fact that it is possible to quantify 
their effect, to provide numbers that can be compared. As mentioned above, the definitions of $\epsilon_{mass}$ and $\epsilon_{env}$ are quite arbitrary in the sense that any author can use their own 
proxy for the environment, which clearly makes comparisons not easy. However, despite that, any author can provide a range of numbers that can quantify the importance of the two modes and their 
dependence on time and on the main physical properties of galaxies such as stellar mass and their environment. Below, we discuss our analysis on $\epsilon_{mass}$ and $\epsilon_{env}$ with a full comparison 
with previous results and, more importantly, their implications. We focus on the dependence of $\epsilon_{env}$ with redshift and environment, of $\epsilon_{mass}$ on redshift and stellar mass,
and separately their mutual dependence, i.e. $\epsilon_{env}$ on stellar mass and $\epsilon_{mass}$ on environment.

\subsection{$\epsilon_{env}$ as a Function of Environment and Redshift}
\label{sec:epsilon_env}
The environmental quenching efficiency has been studied by several authors (e.g. \citealt{vandenbosch08,peng10,quadri12,balogh16,fossati17} and many others), with results that at the time of the writing 
of this manuscript are still controversial (and the same applies for $\epsilon_{mass}$). There is a general agreement that $\epsilon_{env}$ depends on the environment (meant either as clustercentric 
distance, local density or halo mass), but not a full consensus on its dependence on redshift (see e.g., \citealt{quadri12,balogh16} for no dependence and e.g. \citealt{fossati17,kawinwanichakij17,pintoscastro19} for 
dependence). However, there is more evidence that the environmental quenching efficiency depends on redshift. In Figure \ref{fig:fc_redshift} we plot several observed data which, put together, show the 
dependence of $\epsilon_{env}$ on redshift. Despite that, even single studies find a clear trend with redshift. For example, \cite{kawinwanichakij17} explored the evolution of the environmental 
quenching efficiency with redshift for galaxies in different stellar mass bins, finding that it depends both on redshift and stellar mass. Similar conclusions have been reached by \cite{pintoscastro19}, who also 
show $\epsilon_{env}$ vs redshift at different clustercentric distances and find that galaxies closer to the centre have higher quenching efficiencies than those in the field.

Concerning the dependence on the environment, our model completely supports it. Since the environment is usually defined as either the local density (e.g., \citealt{kawinwanichakij17})
or as the clustercentric distance (e.g., \citealt{pintoscastro19}), our analysis extends it to the halo mass. We show in Figure \ref{fig:fc_halomass} that the halo mass acts in a different way to galaxies 
with different stellar mass. Low mass galaxies ($\log M_* <10.5$) suffer from environmental quenching independently of the fact that they reside in a $10^{13} M_{\odot}$ halo or a mature $10^{15} M_{\odot}$
cluster. On the other hand, high mass galaxies are quenched differently depending on the halo mass of the object in which they reside, that is, clusters are more effective than groups or small groups in 
quenching high mass galaxies. Finding the right physical mechanisms responsible for the environmental quenching is beyond the scope of this paper. However, strangulation can be a possible candidate for 
low mass galaxies, since it operates in all haloes independently on their mass, while ram pressure stripping (but a contribution from strangulation can be an option) might be a good candidate for the 
quenching of massive galaxies, since it depends on the mass of the halo and it is stronger in more massive haloes. Moreover, ram pressure stripping is stronger in the central regions of the clusters 
where, for dynamical reasons, more massive galaxies are supposed to fall to in a shorter time.

\subsection{$\epsilon_{mass}$ as a Function of Stellar Mass and Redshift}
\label{sec:epsilon_mass}
The mass quenching efficiency $\epsilon_{mass}$ has been found to be stellar mass and redshift dependent by several authors (\citealt{darvish16,kawinwanichakij17,pintoscastro19}), a result that agrees well 
with the prediction of our model. Most of previous studies find $\epsilon_{mass}$ to increase with cosmic time (as $\epsilon_{env}$), in line with our predictions (see figures \ref{fig:mqe_mass2}, \ref{fig:mqe_halomass} 
and \ref{fig:mqe_redshift}) at least for satellites and in general for massive galaxies. The interesting point to highlight is the different evolution of $\epsilon_{mass}$ with redshift for central and 
satellite galaxies. As pointed out in Section \ref{sec:mass_quenching} and shown in Figure \ref{fig:mqe_redshift}, centrals are quenched by internal processes with roughly the same efficiency, around 0.4, 
regardless the redshift. This means that whatever processes are quenching these galaxies, they do not depend on the cosmic time. Processes such as supernova and AGN feedback can fit in the picture. On the other hand, 
satellites experience different mass quenching efficiencies with time, from around 0.2 at $z\sim 1.5$ to 0.6 at the present time, which means that $\epsilon_{mass}$ is a factor 3 times higher at $z=0$ than it is 
at high redshifts. The obvious conclusion is that for galaxies in dense environments, the internal processes also suffer from the environmental conditions and how they change with time, a hint that $\epsilon_{mass}$ is 
dependent on environment. We will fully discuss this point below. The other important point for what concerns $\epsilon_{mass}$ is its dependence on stellar mass. In agreement with previous work, our model also predicts that 
more massive galaxies are mass quenched more efficiently than less massive ones, independently of their rank, central or satellite, which is in line with the downsizing scenario where more massive galaxies 
quench faster (e.g., \citealt{popesso11,sobral11,fossati17,pintoscastro19}; Rhee J. et al. 2020, in prep).

\subsection{The Mutual Dependence of $\epsilon_{env}$ and $\epsilon_{mass}$}
\label{sec:mutual}
We now move the discussion to the main result of this work, that is the mutual dependence of the two quenching efficiencies. Our analysis shows that the environmental quenching efficiency depends on stellar 
mass and, on the other hand, the mass quenching efficiency depends on the environment, meaning that the two quenching efficiencies are not separable. There is not yet agreement on this important point. Some 
authors find evidence that they are separable, such as \cite{baldry06,vandenbosch08,peng10,quadri12,kovac14,vanderburg18}, and others that they are not, such as \cite{balogh16,darvish16,fossati17,jian17,kawinwanichakij17,papovich18,pintoscastro19}.
Given the importance of the point it is worth discussing the results of a few of the above works. Most of the first studies that focused on the quenching efficiencies argue that the quenching given by the 
environmental processes are not influenced by the stellar mass of the galaxies and, vice versa, that the quenching given by internal processes that are linked to the stellar mass of galaxies are not dependent on 
the particular environment galaxies are living in. This has been shown by \cite{peng10} out to $z\sim 1$, followed by an extension to $z\sim 1.8$ by \cite{quadri12}. \cite{peng10} use a sample of galaxies 
drawn from the Sloan Digital Sky Survey and zCOSMOS and find that the mass quenching efficiency is completely independent of environment meant as local density, and the environmental quenching efficiency is independent of stellar mass. 
\cite{quadri12}, who used a mass-selected sample from UKIDSS Ultra-Deep Survey, echo Peng et al.'s results for what concerns $\epsilon_{env}$ and extend them to higher redshifts. More recently, \cite{vanderburg18}, who 
study a sample of 21 clusters at $z\sim 0.6$ (on average) detected by Planck, find that the environmental quenching efficiency is stellar mass independent in any environment, although it increases toward the inner 
regions of the clusters. 

However, over the last few years, there have been more evidence in support of the fact that the efficiencies of the two quenching modes are not separable. \cite{balogh16} analyse galaxies in groups and clusters at $0.8<z<1.2$
from the GCLASS and GEEC2 surveys and find that $\epsilon_{env}$ (called conversion fraction in their paper) is nearly independent of stellar mass for galaxies more massive than $\log M_* \sim 10.3$, but it decreases 
at lower stellar mass, becoming consistent with zero at $\log M_* \sim 9.5$. The inefficiency of the environment in quenching low mass galaxies has been pointed out also by \cite{papovich18}. As introduced above,
these authors study the evolution of the stellar mass functions of quiescent and star forming galaxies in a wide redshift range and argue that a mass-invariant $\epsilon_{env}$ in the low mass end, below 
$\log M_* \sim 9.5$, should end up in a steeper slope of the stellar mass function of quiescent galaxies than observed. They then conclude that in the range $0.5<z<1.5$, the environmental quenching 
efficiency must decrease with stellar mass. With the same set of data, \cite{kawinwanichakij17} find the same result by looking directly at $\epsilon_{env}$.  Very recently, \cite{pintoscastro19} confirmed all 
these results concerning the mutual dependence of the quenching modes. They take advantage of a sample of more than 200 clusters IR-selected from the ELAIS-N1 and XMM-LSS fields in the redshift range $0.3<z<1.1$. 
By using the clustercentric distance as a proxy of the environment, they find that both efficiencies depend on stellar mass and environment, in good agreement with the above mentioned works and our analysis.

One question arises in very naturally: why the mutual dependence of the two quenching modes is still controversial? Our analysis clearly supports the picture where both efficiencies are environment and stellar 
mass dependent, as did many of the works quoted above. We agree with the argument of \cite{papovich18}, i.e. the reason can be found in the stellar mass limit probed by different observations. For instance, those studies 
that argue for a mass-invariant environmental quenching efficiency (e.g. \citealt{peng10,quadri12,kovac14}) are limited to moderate stellar mass, $\log M_* \gtrsim 10$ in most of the cases. To shed light on this 
point, we need more observations that can probe lower stellar masses, $\log M_* \sim 9-9.5$.

\section{Conclusions}
\label{sec:conclusions}

Taking advantage of an analytic model of galaxy formation that was set to match the evolution of the global stellar mass function from high redshift to the present time and to give, at the same time, good
predictions of the evolution of the SFR$-M_*$ relation, we have investigated the galaxy quenching efficiencies due to environment and stellar mass. These two quenching modes have been analyzed in detail,
by looking at their dependence on redshift, stellar mass and halo mass (which we have used as a proxy for the environment). Given the result outlined from our analysis, we conclude as follows:
\begin{itemize}
  \item The environmental quenching efficiency $\epsilon_{env}$ is a function of halo mass, stellar mass and redshift. The efficiency increases with cosmic time and generally with halo mass. These trends 
        are very neat for galaxies more massive than $\log M_* \sim 10.5$, while for less massive galaxies, we do find an increase with redshift, but a constant relation with halo mass at fixed redshift. 
	$\epsilon_{env}$ depends on stellar mass. At low stellar mass, below $\log M_* \sim 9.5$, the efficiency is consistent with zero. It rapidly increases and reaches the highest values at $\log M_* \sim 10.5$
	(depending on the redshift), to drop again at higher stellar mass. 
  \item The mass quenching efficiency is also a function of stellar mass, halo mass and redshift. For central galaxies, it strongly depends on stellar mass, being much higher for more massive centrals, but very weakly
        on redshift. Similarly, for satellites galaxies, $\epsilon_{mass}$ is in a strong relationship with stellar mass, and also with redshift in the intermediate stellar mass range ($10\lesssim \log M_* \lesssim 11$)
        since $z<1.5$. Moreover, the mass quenching efficiency of galaxies more massive than $\log M_* \sim 10.5$ depends on environment at any redshift, being higher in more massive haloes, while it is constant 
        (with different values depending on the redshift) for less massive galaxies. 
  \item In previous works that studied the two quenching efficiencies, the environment is usually defined as either the local density or as the clustercentric distance. Our analysis extends it to the halo mass, 
        showing that this is also a good proxy for the environment.
  \item The stellar mass and environmental quenching efficiencies are not separable, at any redshift investigated. $\epsilon_{env}$ depends on stellar mass, and vice versa, $\epsilon_{mass}$ (for massive galaxies) depends 
        on the environment. This means that, according to our analysis, intermediate mass galaxies are environmentally quenched faster, and, on the other hand, intermediate/massive galaxies in more massive haloes quench 
        faster. 
  \item Mass quenching mechanisms dominate the quenching of massive galaxies at any redshift, while the environment becomes gradually more important as time goes by in the intermediate stellar mass range and dominates
        at lower stellar masses, $\log M_* <10$. At stellar masses lower than $\log M_* \lesssim 9.5$ both quenching mechanisms become inefficient, regardless of the redshift.
\end{itemize}
The predictions of our model agree qualitatively well with the results of most of the studies quoted above. The general picture supported and proved by this analysis sees the two quenching modes to be both stellar 
mass, environment and redshift dependent. In the context of "nature" versus "nurture``, these results prove that they are both important for galaxy evolution, interconnected in a nontrivial way. The redshift, 
stellar mass and halo mass dependences of both quenching modes for galaxies in groups and clusters are particularly interesting because they highlight the need to invoke a plethora of physical processes that act 
with different timescales, at different stellar mass and halo mass scales. Starvation and ram pressure stripping are good candidates for the environment-driven processes, while, except for AGN and supernova feedback, 
most of the difference in the quenching due to internal processes can arise from the fraction pf stellar/gas mass that galaxies have at the time of accretion. 

\section*{Acknowledgements}
The authors thank the anonymous referee for his/her constructive comments which helped to improve the manuscript.
This work is supported by the National Key Research and Development Program of China (No. 2017YFA0402703), 
by the National Natural Science Foundation of China (Key Project No. 11733002), and the NSFC grant 
(11825303, 11861131006). S.K.Y. acknowledges support from the Korean National Research Foundation 
(NRF-2017R1A2A05001116).

\label{lastpage}

\end{document}